\documentclass[%
 aip,
 amsmath,amssymb,floatfix,
 reprint,%
]{revtex4-1}
\usepackage[breaklinks=true,colorlinks=true,linkcolor=blue,urlcolor=blue,citecolor=blue]{hyperref}
\usepackage{graphicx}
\usepackage{dcolumn}
\usepackage{bm}
\usepackage{changes}

\usepackage[utf8]{inputenc}
\usepackage{mathptmx}
\usepackage{etoolbox}
\usepackage{booktabs}
\usepackage{physics}
\usepackage{tabularx} 
\usepackage{amssymb}
\usepackage{amsmath}
\usepackage[]{units}
\usepackage[]{upgreek}
\UseRawInputEncoding 
\makeatletter
\def\@email#1#2{%
 \endgroup
 \patchcmd{\titleblock@produce}
  {\frontmatter@RRAPformat}
  {\frontmatter@RRAPformat{\produce@RRAP{*#1\href{mailto:#2}{#2}}}\frontmatter@RRAPformat}
  {}{}
}%
\makeatother
\begin{document}

\preprint{AIP/123-QED}

\title[]{Hybrid Magnonic-Oscillator System}
\author{A. Hamadeh}
 \email{hamadeh@rhrk-uni-kl.de}
\affiliation{Fachbereich Physik and Landesforschungszentrum OPTIMAS, Technische Universit\"at Kaiserslautern, 67663 Kaiserslautern, Germany}
\author{D. Breitbach }%
\affiliation{Fachbereich Physik and Landesforschungszentrum OPTIMAS, Technische Universit\"at Kaiserslautern, 67663 Kaiserslautern, Germany}
\author{M. Ender }%
\affiliation{Fachbereich Physik and Landesforschungszentrum OPTIMAS, Technische Universit\"at Kaiserslautern, 67663 Kaiserslautern, Germany}
\author{A. Koujok}%
\affiliation{Fachbereich Physik and Landesforschungszentrum OPTIMAS, Technische Universit\"at Kaiserslautern, 67663 Kaiserslautern, Germany}
\author{M. Mohseni}%
\affiliation{Fachbereich Physik and Landesforschungszentrum OPTIMAS, Technische Universit\"at Kaiserslautern, 67663 Kaiserslautern, Germany}
\author{F. Kohl}%
\affiliation{Fachbereich Physik and Landesforschungszentrum OPTIMAS, Technische Universit\"at Kaiserslautern, 67663 Kaiserslautern, Germany}
\author{J. Maskill}%
\affiliation{Fachbereich Physik and Landesforschungszentrum OPTIMAS, Technische Universit\"at Kaiserslautern, 67663 Kaiserslautern, Germany}
\author{M. Bechberger}%
\affiliation{Fachbereich Physik and Landesforschungszentrum OPTIMAS, Technische Universit\"at Kaiserslautern, 67663 Kaiserslautern, Germany}
\author{P. Pirro}%
\email{ppirro@rhrk-uni-kl.de}
\affiliation{Fachbereich Physik and Landesforschungszentrum OPTIMAS, Technische Universit\"at Kaiserslautern, 67663 Kaiserslautern, Germany}
\date{\today}

\begin{abstract}

We propose a hybrid magnonic-oscillator system based on the combination of a spin transfer auto-oscillator and a magnonic waveguide to open new perspectives for spin-wave based circuits. The system is composed of a spin transfer oscillator based on a vortex state which is dipolarly coupled to a nanoscale spin-wave waveguide with longitudinal magnetization. In its auto-oscillating regime, the oscillator emits coherent spin waves with tunable and controllable frequencies, directions and amplitudes into the waveguide. We demonstrate the principle of this method using micromagnetic simulations and show that reconfiguration of the system is possible by changing the chirality and polarity of the magnetic vortex. Spin waves are emitted into the waveguide with high non-reciprocity and the preferred direction depends on the core polarity of the vortex. In contrast, different vortex chiralities lead to different amplitudes of the emitted waves. Our findings open up a novel way 
to design an agile spintronic device for the coherent and tunable generation of propagating spin waves.


\end{abstract}

\maketitle

\section{\label{sec:level1} Introduction}

The investigation of spin waves goes a long way back to the twentieth century \cite{bloch1930theorie,van1958spin}. Since then, research in these fundamental magnetic excitations has been gaining increasing attention. Spin waves are studied due to fundamental interest  \cite{Serga_2010,Kruglyak_2010} , as well as for their large potential for information processing applications, such as boolean spin-wave logic devices or alternative computing schemes \cite{kostylev2005spin, vogt2014realization, chumak2015magnon, yu2013omnidirectional}. Perhaps some of the most important advantages of spin waves are their ability to transport information without generating ohmic losses \cite{kruglyak2006magnonics,kruglyak2010magnonics,lenk2011building}, as well as the possibility of being used in nanoscale spintronic signal-processing devices \cite{demidov2010direct,madami2011direct,yu2013omnidirectional,vogt2014realization}. When it comes to the realization of spin-wave devices, some of the most important ingredients are ways to efficiently excite spin-waves. The most common method for this task is direct spin-wave excitation by microwave currents in antennas \cite{demokritov2006bose, buttner2000linear, park2002spatially, covington2002time}. Further methods involve the utilization of the spin-transfer torque (STT) \cite{slonczewski1996current,berger1996emission,kiselev2003microwave}  or ultrashort light pulses \cite{van2002all,kimel2005ultrafast,satoh2012directional}. Another developed method of spin-wave excitation is  proposed to use the driven rotational dynamics of a formed magnetic vortex cores in an antiferro - magnetically coupled trilayer \cite{wintz2016magnetic}. The vortex configuration is a region of in-plane curling magnetization at the edges, and an out-of-plane magnetization at the center. It is characterized by two degrees of freedom, namely the polarity (P) and chirality (C). The polarity describes the direction of the out-of-plane magnetization at the vortex core, with positive/negative polarities referring to upward/downward directions of magnetization. The chirality describes the direction of the in-plane magnetization, with positive/negative chiralities referring to clockwise/counter clockwise directions of magnetization \cite{shinjo2000magnetic}. When STT is used to drive the auto-oscillation of the magnetic vortex, we refer to these devices as spin-transfer torque vortex oscillators (STVOs) \cite{pribiag2007magnetic,dussaux2010large}. STVOs are magnetic oscillators in the GHz range which are driven by a DC current and thus constitute very promising and energy efficient sources for GHz excitations \cite{villard2009ghz,zeng2013spin,kim2012spin}. Due to their intrinsic non-linearity and their ability to synchronize their oscillation frequencies, STVOs are already used in novel logic concepts such as  brain inspired and neuromorphic computing \cite{locatelli2014spin, torrejon2017neuromorphic, grollier2020neuromorphic,hamadeh2014perfect} or vowel recognition \cite{romera2018vowel}. On the other hand, magnons, the eigen-excitations of the spin system, are proposed as data carriers for future beyond-CMOS logic. Several logic building blocks such as magnon transistors \cite{chumak2014magnon,wright2018trio}, majority gates \cite{klingler2014design} or nonlinear directional magnon couplers \cite{wang2019realization} have been realised. Based on this, we propose to merge these two important sub-fields of spintronics, namely the field of spin-transfer torque vortex oscillators (STVOs) \cite{csaba2020coupled} and the field of magnonics\cite{khitun2010magnonic,wang2020magnonic,pirro2021advances}. Here we demonstrate a novel hybrid device concept which unifies the particular strength of both fields, which we refer to as the hybrid magnonic-oscillator system (HMOS). 
The proposed device consists of a vortex-based spin-transfer torque oscillator, which is dipolarly coupled to an adjacent magnonic waveguide. A coupled system where spin waves propagating in a waveguide are coupled to spin-wave modes in a vortex state has been previously studied in \cite{korber2020nonlocal}. In this case however, no auto-oscillation and vortex core dynamics have been involved. Here, we numerically study this system using the open source micromagnetic solver MuMax 3.0 package \cite{doi:10.1063/1.4899186}, as well as the software platform Aithericon \cite{Aithericon}. We demonstrate that, when carefully designed, this hybrid system can be used as a tunable source for the emission for coherent spin waves in the magnonic waveguide. In the following, we investigate the spin-wave dispersion relation in the bare magnonic waveguide. We then numerically investigate the coupled HMOS system and the emission of spin-waves by the vortex oscillator. In particular, we study the effects of changing the polarity (P) and chirality (C) of the vortex on the emitted signal intensity and directionality. Finally, we study the evolution of the spin-wave frequency, intensity and non-reciprocity as a function of the current density.

\section{SIMULATION SETUP}

\begin{figure}[!ht]
  \includegraphics[width=0.99\columnwidth]{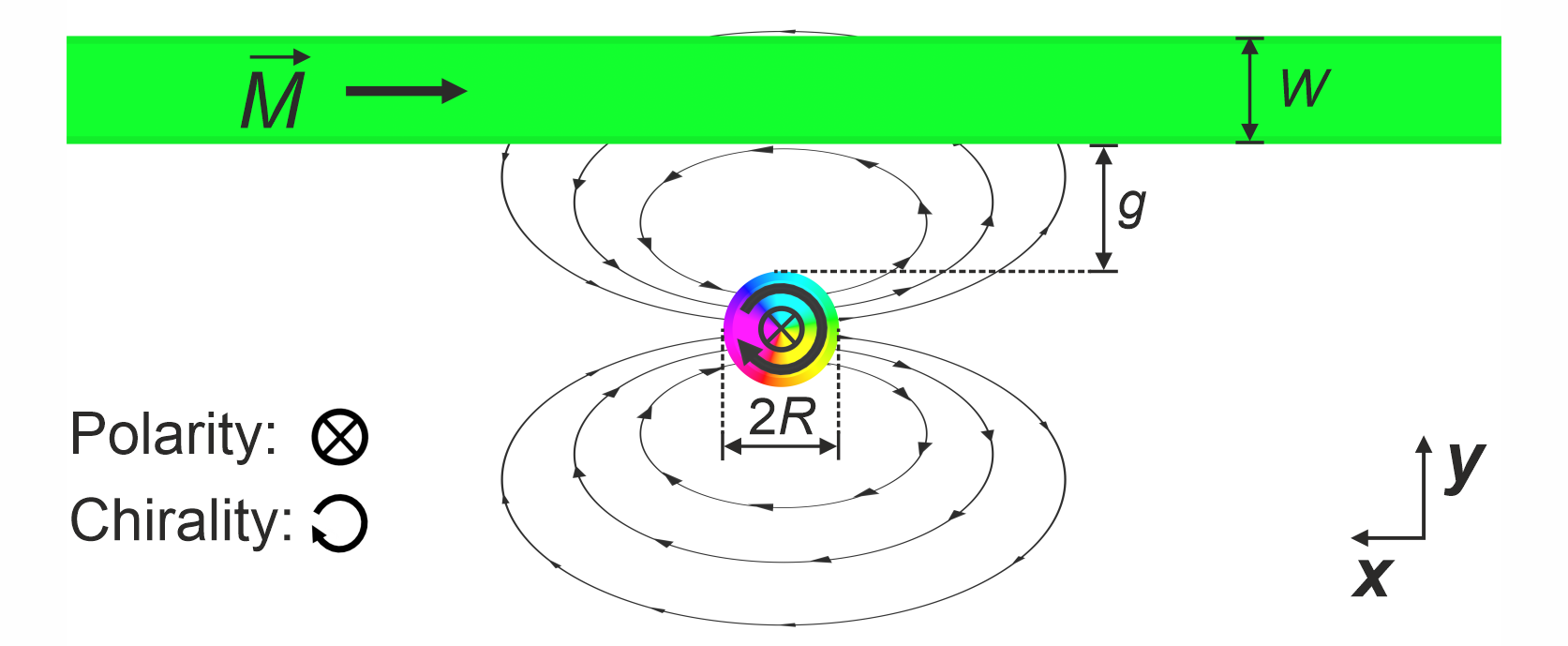}
 \caption{Schematic representation of the hybrid magnonic-oscillator system (HMOS). The uniform magnetization $\Vec{M}$ along the waveguide is pointing to the right. As an example, the vortex is shown here with clockwise chirality and negative polarity. To illustrate the dipolar coupling, the dipolar stray fields are shown as emanating from a magnetic dipole at the position of the oscillator.}
  \label{fig:1}
\end{figure}
The proposed hybrid magnonic-oscillator system is schematically depicted in FIG. \ref{fig:1}. It consists of a magnonic waveguide made of Yttrium Iron Garnet (YIG) with a uniform magnetization $\Vec{M}$ directed along its long axis (x-direction) and in the absence of an external magnetic bias field. Adjacent to the waveguide is a Cobalt-Iron-Boron (CoFeB) disk, operating as a spin-transfer vortex oscillator (STVO), which can be driven to auto-oscillation if an electric current is passed through it. The eigen-frequencies of the two sub-systems can be designed to match the desired functionality of the hybrid structure.
\noindent
On the one hand, the gyrotropic frequency of the vortex oscillation ($\omega_\text{v}$) for thin disks in the absence of an external magnetic field depends on the thickness ($t_{\mathrm{disk}}$), the radius ($R$) of the disk, as well as on the saturation magnetization $M_\text{s}$ \cite{guslienko2002eigenfrequencies}. For $\frac{t_{\mathrm{disk}}}{R}<<1$, $\omega_\text{v}$ is approximated as:

\begin{equation*}
\left\{
\omega_\text{v}\approx\frac{20}{9}\gamma M_\text{s}\frac{t_{\mathrm{disk}}}{R}
\right\}.
\end{equation*}


\noindent On the other hand, the dispersion relation for spin waves in this waveguide geometry is given by \cite{Wang2019_PRL}:\\

\begin{center}
    $\omega(k_x) = \sqrt{(\omega_\text{H}+\omega_\text{M}(\lambda^2 K^2 + F_{k_x}^{yy}))
    (\omega_\text{H}+\omega_\text{M}(\lambda^2 K^2 + F_{k_x}^{zz}))}$,
    
\end{center}
\vspace{0.2 cm}
\noindent where $\omega_\text{H}$ = $\gamma\mathrm{B}$, $\omega_\text{M}$=  $\gamma \mu_\text{0} M_\text{s}$ , $\lambda =\sqrt{\frac{2A}{M^2_\text{s}\mu_\text{0}}}$ is the exchange length, $A$ is the exchange constant, $M_\text{s}$ is the saturation magnetization, $K = \sqrt{k_x^2+\kappa^2}$ is the total in-plane wave number with its component $k_x$ along the propagation direction and $\kappa$ is the effective wave number across the waveguide width. The tensor $F_{k_x}$ describes the dynamical magneto-dipolar interaction, $\mathrm{B}$ is the static internal magnetic field, $\mu_\text{0}$ is the vacuum permeability and $\gamma$ is the gyromagnetic ratio. \\
Thus, besides selecting  particular materials for the oscillator and the waveguide, one can tune the individual resonance frequencies by changing the thickness and width of the waveguide, as well as the diameter and thickness of the disk to reach a resonant coupling between the STVO and the magnonic waveguide. To ensure realistic conditions, we have used standard parameters of nanometer-thick YIG films \cite{dubs2020low}, and circular nanopillars with diameters of \unit[100]{nm} for the oscillator. At the oscillator, the magnetic configuration of the free layer is a vortex, with the possibility to tune the core polarities and chiralities \cite{jenkins2014controlling,locatelli2011dynamics}. We focus here on the gyrotropric  oscillation mode because the gyration of a vortex core through spin-torque in a single pillar is experimentally well proven and understood \cite{hamadeh2014origin}. Analytical descriptions of these dynamics have been shown to be quantitatively consistent with experiments \cite{dussaux2012field}. In addition, vortex oscillators exhibit low phase noise and have been experimentally shown to synchronize through dipolar coupling \cite{locatelli2015efficient}. In the numerical simulations, we assume that the magnetization of the polarizing magnetic layer is fixed and points out-of-plane and that the magneto-static field emanating from it is negligible. The magnetic parameters used for the simulations are displayed in Table \ref{Table_p3460}.

\begin{table}[!ht]
  \begin{tabularx}{0.4\textwidth}{XXXXX}
      \hline\hline \\
      Parameter &  Disk  & Waveguide\\
      \hline \\
      $M_\text{sat}$ (A/m) & 1.4$\times$10$^{6}$ & 1.4$\times$10$^{5}$\\
      \hline \\
      $A_\text{ex}$ (J/m) & 1.6$\times$10$^{-11}$ & 3.5$\times$10$^{-12}$\\
      \hline \\
      $\alpha$ & 2$\times$10$^{-3}$ & 2$\times$10$^{-4}$\\
      \hline\hline
  \end{tabularx}
  \caption{The parameters for micromagnetic modeling of the HMOS in MuMax$^3$.}
   \label{Table_p3460} 
\end{table}

In the waveguide the Gilbert damping parameter $\alpha$ is set to a value of \unit[2$\times$10$^{-4}$]{}. In addition, to avoid spin-wave reflections, absorbing boundary conditions were modeled by exponentially increasing $\alpha$ at the ends of the waveguide to a value of \unit[0.5]{}. The mesh size was set to $5 \times 5 \times$\unit[10]{nm}$^3$ (two layers in z-direction). The waveguide has a total length of $\unit[20.48]{\upmu m}$, a width of $W = \unit[100]{nm}$ and a thickness of $t_{\mathrm{waveguide}} = \unit[20]{nm}$. The oscillator was placed in a distance of $g=\unit[120]{nm}$ from the waveguide. The oscillator has a thickness of $t_{\mathrm{disk}} = \unit[20]{nm}$ and a disk diameter of $2R = \unit[100]{nm}$. To excite the vortex oscillation via the spin-transfer torque effect, a DC current injected through the disk in out-of-plane direction is modeled (Simulation parameters: Spin polarization $P$= 0.5, degree of non-adiabaticity $\xi$ = 0, Slonczewski secondary STT term = 0). We note that in the simulation, only this spin-polarized current injected into the free layer is explicitly modeled. In the experiment, such a current is generated, for example, by a fixed magnetic layer, which is not explicitly modeled in our simulations. Also the Oersted field generated by the charge current is computed analytically and plugged into Mumax$^3$. The y-component of the magnetization $M_{\mathrm{y}}(x,y,t)$ of each cell was collected over a time period of \unit[200]{ns} and recorded in intervals of \unit[0.08]{ns}. The spin-wave dispersion was calculated by performing a fast Fourier transform (FFT) in space and time of $M_{\mathrm{y}}$ in the waveguide, and the frequency of the oscillator is extracted from the fast Fourier transform of $M_{\mathrm{y}}$ in the disk. 
Here it is worth noting that we chose the parameters of CoFeB for the oscillator since this material has a high saturation magnetization $M_\mathrm{s}$ which increases the frequency of the oscillator pushing it closer to the spin-wave dispersion in the YIG waveguide.

\section{RESULTS AND DISCUSSION}

\begin{figure}[!ht]
 \includegraphics[width=\columnwidth]{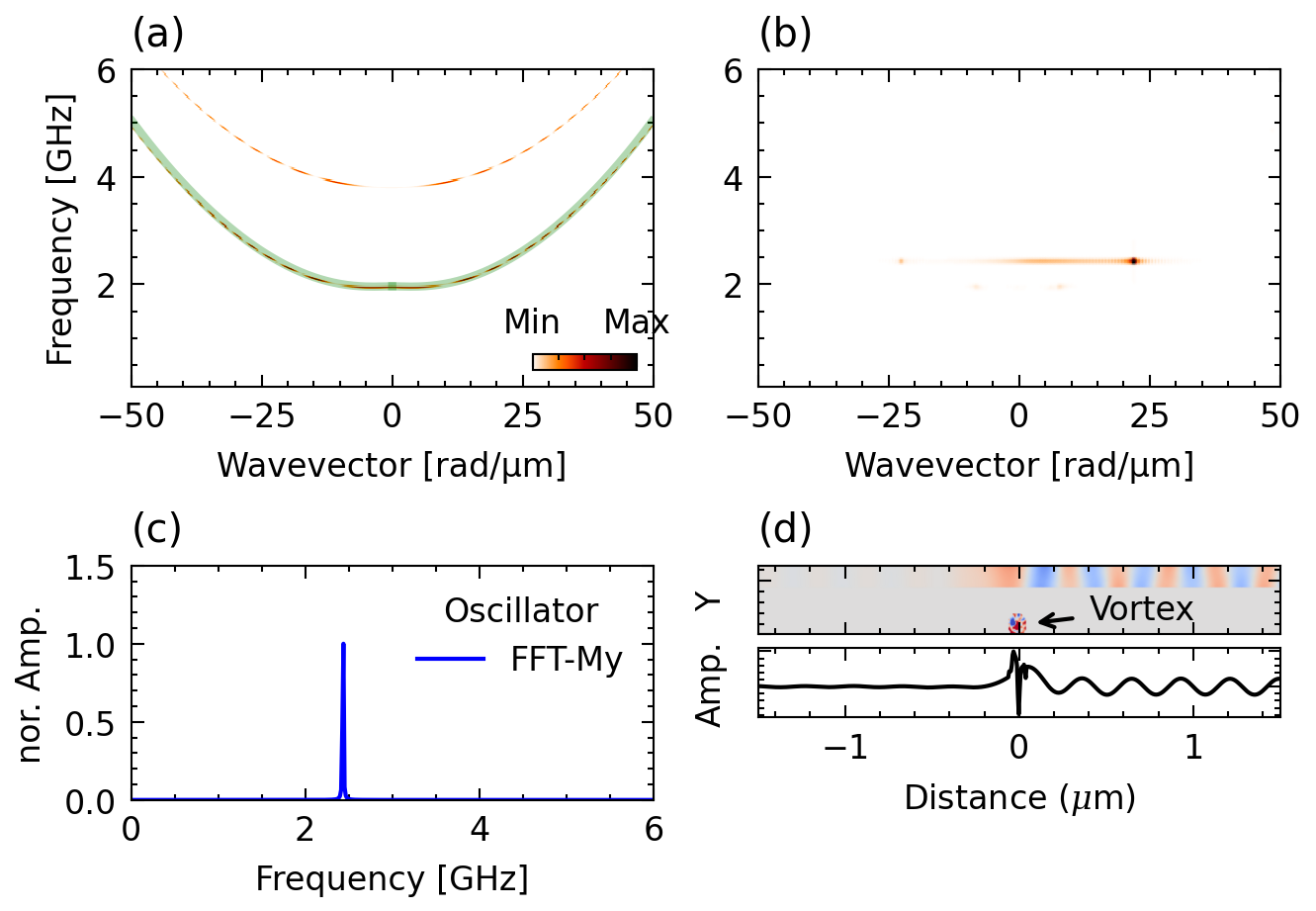}
 \caption{Dispersion relation of spin waves propagating in the waveguide (a) Spin-wave spectrum excited by an antenna by applying a sinc-function field at the middle of the waveguide. The two lowest waveguide modes are visible. The green line shows the analytical dispersion relation for the uniform mode which has been calculated using the same parameters as in the simulation \cite{wang2018reconfigurable} showing excellent agreement.(b) Spin waves excited by the spin transfer oscillator based on the vortex state (the electrical current density is $J_{\mathrm{DC}}$=\unit[1.2$\times10^{11}$]{$A/m^2$}. 
 One recognizes a dominant excitation of spin waves with $k_x \approx 25$ rad$/ \upmu$m propagating to the right at a frequency of $f_\mathrm{SW}=\unit[2.5]{GHz}$. (c) FFT spectra of $M_\text{y}$ of the oscillator showing that $f_{\mathrm{Osci}}$ = $\unit[2.5]{GHz}=f_\mathrm{SW}$. (d) Snapshots of the spatial profile of $M_\text{z}$ (upper subfigure) and $M_{\mathrm{y}}$ integrated over the waveguide width (lower subfigure), as a function of the distance x from the oscillator along the waveguide, respectively.}
 \label{fig:2}
\end{figure}

We illustrate the waveguide dispersion relation using broadband excitation by means of an antenna in the center of the waveguide. The resulting mode spectrum of the waveguide is shown in FIG. \ref{fig:2} (a). It contains two spin-wave modes. The low frequency mode is the fundamental waveguide mode with uniform oscillation across the width, while the higher frequency mode is the first width mode of the waveguide. This result reflects a typical waveguide dispersion curve for a width confined nanostripe geometry \cite{Wang2019_PRL}. For comparison, the fundamental mode was calculated analytically (green curve). FIG. \ref{fig:2} (b) shows the spin-wave spectrum in the waveguide, when the vortex oscillator is driven to auto-oscillation by a constantly injected spin current. The vortex oscillation couples dipolarly to the waveguide and excites spin waves at a frequency of $f_{\mathrm{Emission}}=\unit[2.5]{GHz}$. FIG. \ref{fig:2} (c) shows the FFT of the component $M_\text{y}$ of the disk, showing that the auto-oscillation frequency of $f_{\mathrm{Osci}}=\unit[2.5]{GHz}$ is equal to the frequency of the emitted spin waves. Adjusting the current allows this frequency to be changed, as will be shown later.

FIG. \ref{fig:2} (d) shows the propagation of spin waves along the waveguide in the respective positive direction (to the right), as well as the distribution of the magnetization $M_\text{y}$ along the longitudinal axis. The dominant propagation of spin waves in the positive direction as shown in FIG. \ref{fig:2} for the studied vortex chirality and waveguide magnetization introduced in FIG. \ref{fig:1} indicates that the emission is unidirectional and that it is possible to control the generated spin waves by simply changing the two degrees of freedom of the vortex, i.e., its polarity and chirality.

\begin{figure*}
\includegraphics[width=2\columnwidth]{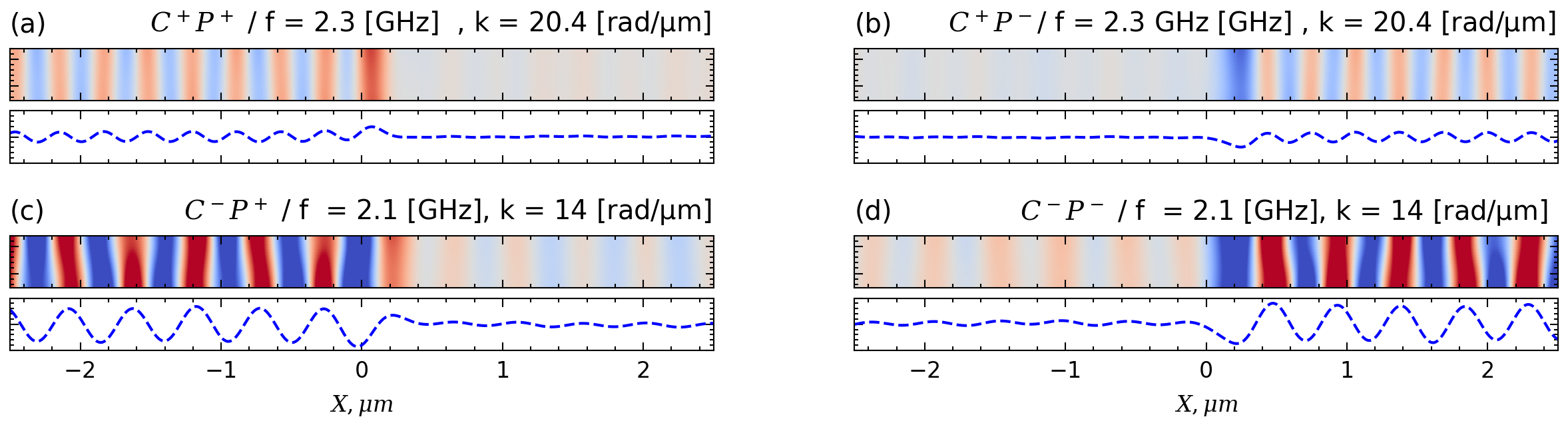}
\caption{\label{fig:wide} Two-dimensional maps of spin-wave emission showing different vortex polarity and chirality combinations and their respective effect on the direction and amplitude of the emitted spin-waves. Four different cases of different polarity and chirality combinations were presented by combining the polarities P (P$^+$: upward (a,c) and P$^-$: downward (b,d)) and the chiralities (C$^+$: clockwise and C$^-$: counter clockwise). The figure shows the out-of-plane spatial distributions of $M_{\mathrm{z}}$ and line profiles represent $M_{\mathrm{y}}$  integrated over the waveguide width  along the longitudinal axis (lower sub-figure) , as obtained from simulation for fixed times. The injected  current density is  $J$ =  \unit[5$\times10^{10}$]{$A/m^2$} for P$^+$ and $J$ =  - \unit[5$\times10^{10}$]{$A/m^2$} for P$^-$.}
\end{figure*}

\begin{figure}
\includegraphics[width=\columnwidth]{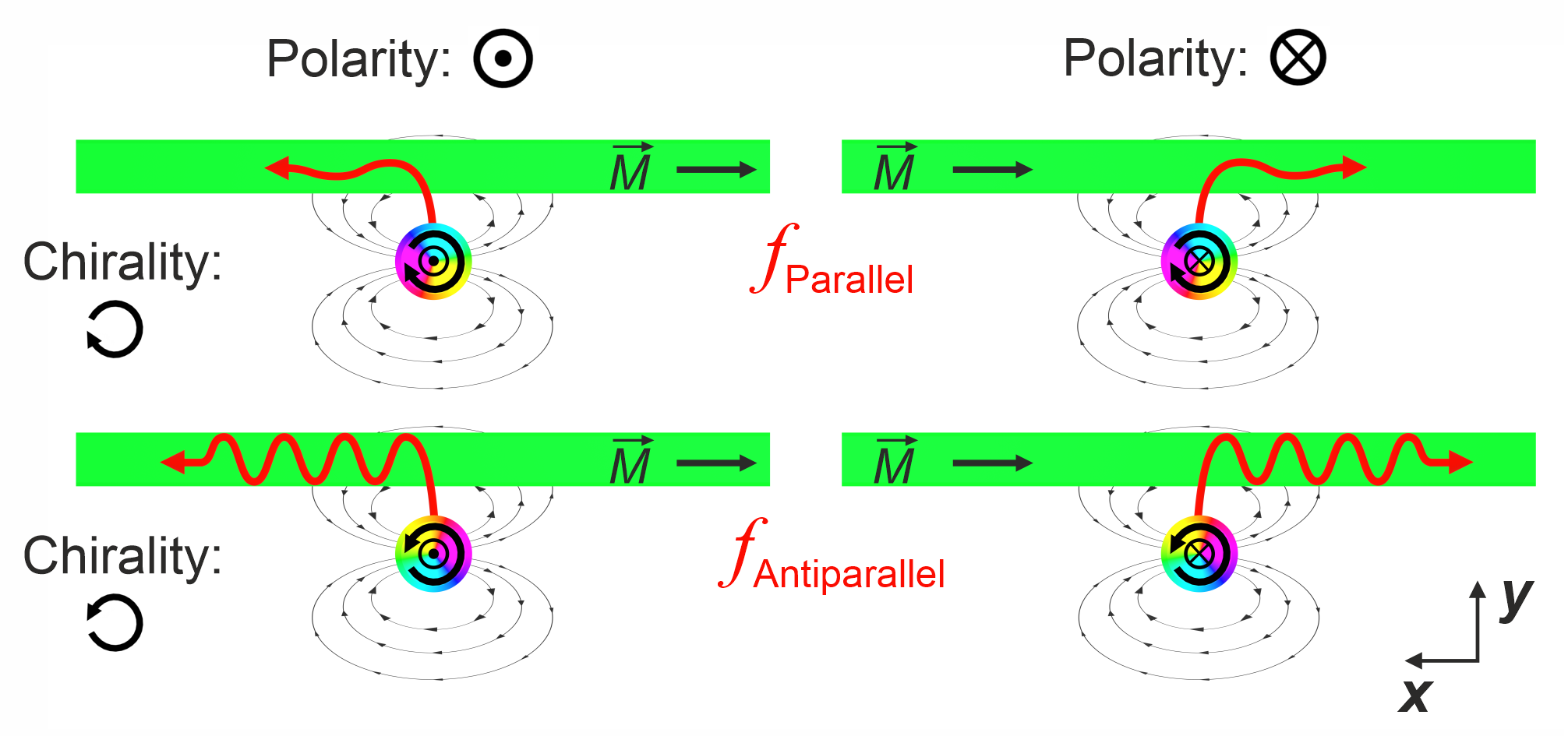}
\caption{\label{fig:4} Schematic representation of the HMOS emission behaviour, based on the results of the micromagnetic simulations. While the polarity P controls the emission direction, the chirality C defines amplitude and frequency of the emitted spin waves.}
\end{figure}

In order to demonstrate the reconfigurability of the emission direction and amplitude, we present results of micromagnetic simulations for different vortex configurations in FIG. \ref{fig:wide} for a fixed injected charge current density of $ \mid J \mid$ =  \unit[5$\times$10$^{10}$]{$A/m^2$}.  All four different combinations of vortex chirality/polarity are displayed. FIG. \ref{fig:wide} (a) refers to the case of a vortex with clockwise chirality $\circlearrowright$   and a positive polarity (C$^+$P$^+$). In this case, the vortex core gyrates counterclockwise $\circlearrowleft$. The generated spin waves propagate to the left with a relatively small amplitude. When changing only the polarity (C$^+$P$^-$) and the direction of the current (required to keep the vortex auto-oscillation with STT), the vortex core gyrates clockwise $\circlearrowright$ and the spin waves propagate to the right with an amplitude that is nearly equal to that of case (a) (see  FIG. \ref{fig:wide} (b)). In both cases, the spin waves have a frequency of $f=\unit[2.3]{GHz}$, and a wave vector $k=\unit[20.4]{rad/\upmu m}$. When changing the chirality, as displayed in (c) (C$^-$P$^+$) and (d) (C$^-$P$^-$), the emitted spin waves have a larger amplitude compared to cases (a) and (b), yet the directional dependence on the polarity is the same. However, the frequency and wave vector of the emitted spin waves in the cases (c) and (d) are different $f = \unit[2.1]{GHz}$ and $k = \unit[14]{rad/\upmu m}$ \cite{jenkins2014controlling}. It can thus be noted that changing the vortex polarity leads to changing the direction of the emitted spin waves along the waveguide. On the other side, the direction of the chirality relative to the waveguide magnetization affects the amplitude of the generated spin waves. 

A schematic representation demonstrating the directions and amplitudes of the emitted spin waves throughout the HMOS is presented in FIG. \ref{fig:4}. The effect of the vortex chirality can be understood by considering the characteristics of dipolarly coupled magnonic nanostructures. Different chiralities lead to an anti-parallel or parallel orientation of the local magnetization of the oscillator and the waveguide. Since the dipolar coupling is stronger for the anti-parallel alignment \cite{wang2018reconfigurable}, the emitted spin-wave amplitude is larger in the presented case for a counter-clockwise vortex chirality. The situation is inveresed if the static magnetisation in the waveguide is inversed.

\begin{figure}[!ht]
  \includegraphics[width=\columnwidth]{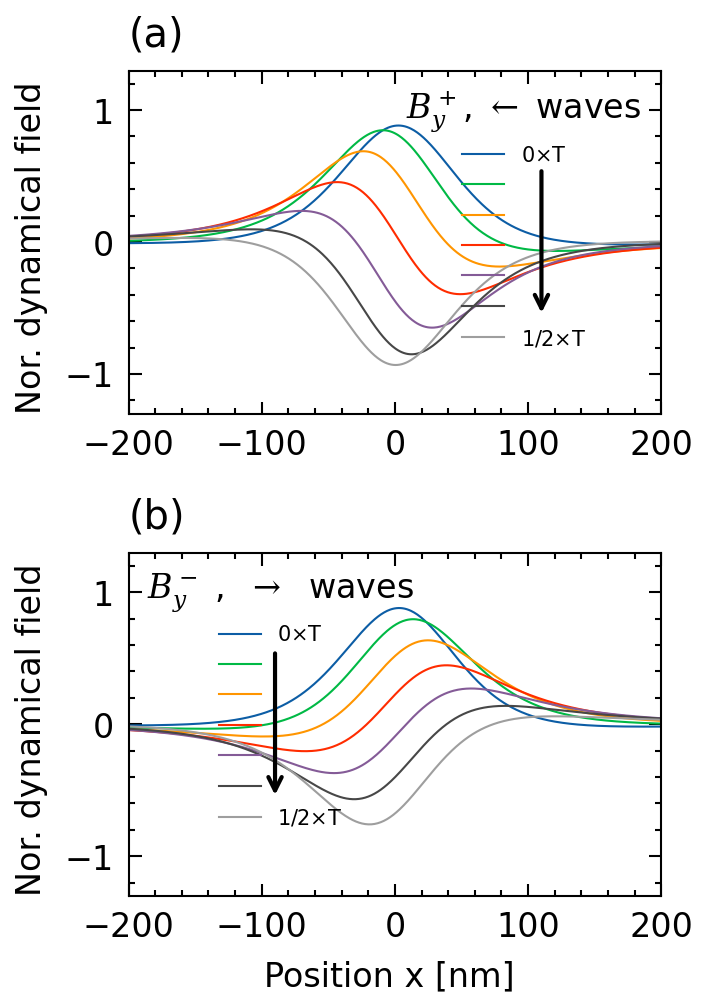}
 \caption{Dynamic magnetic field generated by a vortex at the location of the waveguide. The dynamic field's strength $B_\text{y}$ is presented for the two cases (a) vortex with positive polarity, (b) vortex with negative polarity as a function of its waveguide position for different oscillation periods ($0T\longrightarrow\frac{1}{2}T$). The respective generated spin waves traverse the waveguide in opposite directions for  $B^+_\text{y}$/$B^-_\text{y}$. For more clarity see the attached animation Fig5annimation.mp4.}
  \label{fig:5}
\end{figure}

A noticeable characteristic of the generated spin waves is their unidirectionality which appears despite the total reciprocity of the spin-wave dispersion for the presented waveguide geometry \cite{Mohseni2019_PRL}. As seen in FIG. \ref{fig:wide}, the generated spin waves propagate in the waveguide in a single direction with maximum amplitude, and an almost vanishing amplitude in the opposite direction. To illustrate the origin of this complex behavior, we performed a vortex-only simulation to obtain the dynamic fields generated by the rotating vortex at the position of the waveguide. In FIG. \ref{fig:5}, we present the amplitude of $B_\text{y}$, which is the main component of the dynamic dipolar field exerting a torque on the wave guide's magnetization, as a function of its position along the waveguide. Displayed is the evolution of $B_\text{y}$ for half of a vortex gyration cycle $T$ ($0T$ to $\frac{1}{2}$T). FIG. \ref{fig:5} (a) refers to the case of the vortex with positive polarity $P^+$ which results in a counter-clockwise vortex core gyration $\circlearrowleft$ (the sense of the vortex core gyration must not be confused with the chirality C of the static vortex) . In this case, the spin waves are emitted into the waveguide in the negative direction of the x-axis $\longleftarrow$ (refer back to FIG. \ref{fig:wide}).  As can be seen from FIG. \ref{fig:5} (a), the dynamic dipolar field  $B^+_\text{y}$ is not similar to a standing wave but it has characteristics of a wave moving from right to left. This can be understood considering the direction of the vortex core movement when the latter is closest to the guideline in its upper half of the oscillation circle. Consequently, the direction of the vortex core movement when close to guideline also agrees with the direction of the spin-wave emission.
 FIG. \ref{fig:5} (b) presents the alternative case of a vortex with negative polarity $P^-$ where the core's gyration trajectory is clockwise  $\circlearrowright$. Here, the spin waves are emitted into the waveguide in the positive direction of the x-axis (refer back to FIG. \ref{fig:wide}). Due to the opposite sense of the vortex core gyration, the dipolar field $B^-_\text{y}$ has characteristics of a wave moving from left to right ($\longrightarrow$) which again agrees with the vortex core movement when close to guideline. 

\begin{figure}[!ht]
  \includegraphics[width=\columnwidth]{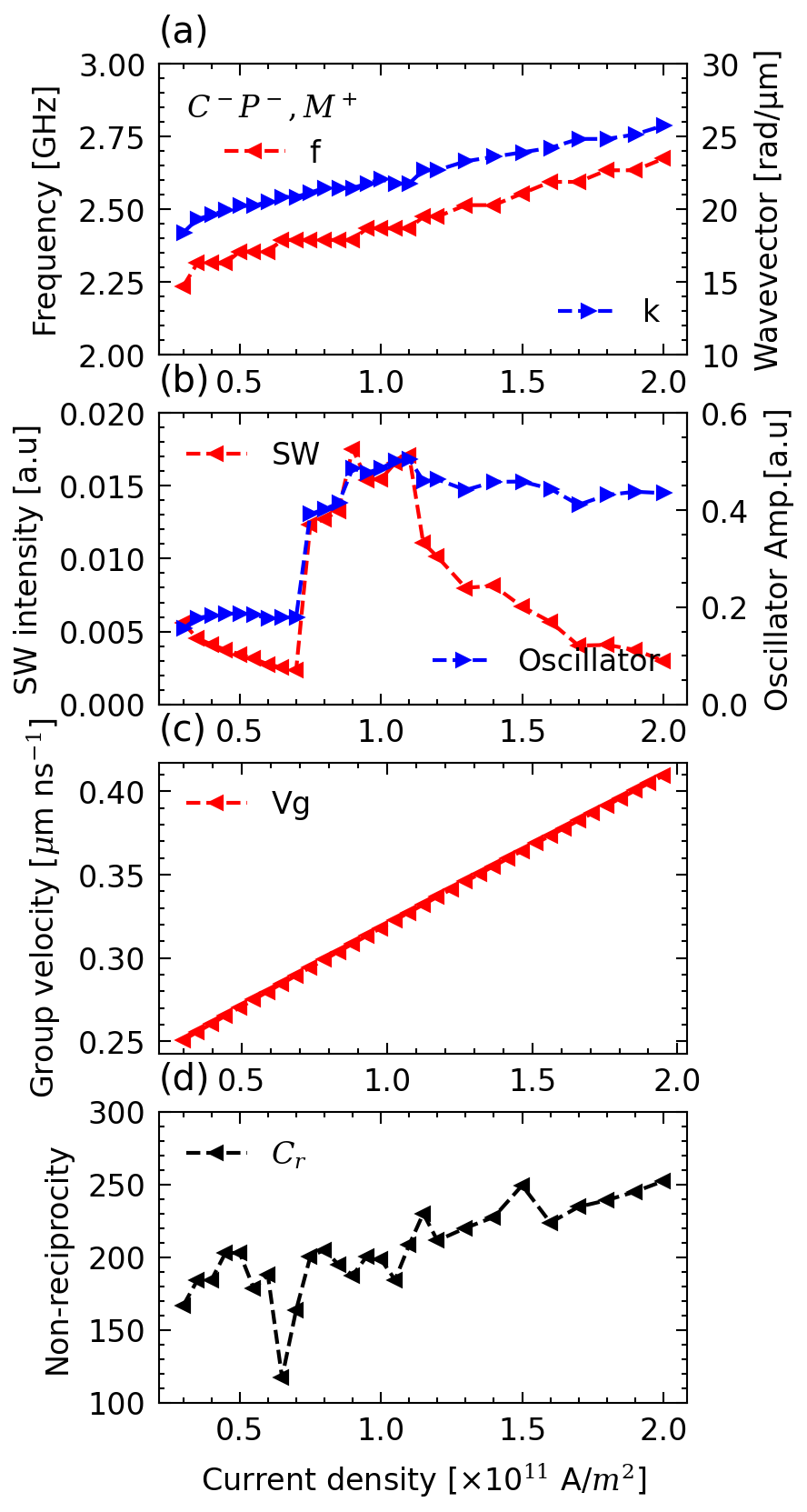}
 \caption{Characteristics of spin-wave emission as a function of current density $J$ (in A/$m^2$) injected in the oscillator for the case of a vortex with negative polarity and chirality (C$^-$P$^-$). (a) Evolution of the frequencies $f$ in GHz (red) and wave vector in rad/$\upmu$m (blue) as a function of $J$. (b) evolution of the spin-wave intensity (red) and oscillator amplitude (blue) as a function of $J$. (c) The spin-wave group velocity $v_{\mathrm{g}}$ as a function of $J$. (e) Non-reciprocity $C_{\mathrm{r}}$ as a function of $J$.}
  \label{fig:6}
\end{figure}

Finally, we investigate the spin waves' characteristics as a function of the current density $J$ (in A/$m^2$) (see FIG. \ref{fig:6})) for $M^+$ (magnetization pointing along the positive axis of the waveguide. The vortex polarity and chirality are set as (C$^-$P$^-$) to ensure strong spin-wave emission into positive x-direction. In FIG. \ref{fig:6} (a), the evolution  of the frequency $f$ and wave vector $k$ are shown to increase from $f=\unit[2.25]{GHz}$ to $f=\unit[2.65]{GHz}$ and from $k=\unit[20]{rad/\upmu m}$ to $k=\unit[26]{rad/\upmu m}$ respectively, as a function of $J$ ranging from \unit[0.2$\times$10$^{11}$]{$A/m^2$} to \unit[2$\times$10$^{11}$]{$A/m^2$}. This well-known increase in the system's frequency is accompanied by an initial increase in the oscillator's amplitude as well as an increase in the intensity of the generated spin waves (see FIG. \ref{fig:6} (b)). With the steady increase in frequency, the vortex gradually reaches a state of nearly constant amplitude at a frequency around \unit[2.35]{GHz} (current density of \unit[0.75$\times$10$^{11}$]{$A/m^2$} ). Beyond this point, the reached amplitude is then maintained by the oscillator despite the increasing frequency and current density since the gyration of the vortex has reached the largest possible radius and a further increase is suppressed by nonlinear effects . In this case, the increase in frequency translates to an increase in the vortex core's gyration speed \cite{guslienko2008dynamic}. Interestingly, the intensity of the generated spin waves drops at the time the oscillator reaches a constant amplitude for frequencies exceeding \unit[2.45]{GHz}. The energy flow carried by the waves is proportional to the product of intensity and the wave group velocity. Concerning the emitted energy, the drop in intensity  is at least partially compensated by an increase in the group velocity $v_{\mathrm{g}}$ of the spin waves (see FIG. \ref{fig:6} (c)). FIG. \ref{fig:6} (d) is a plot of the spin waves' non-reciprocity C$_\text{r}$ as a function of $J$, where C$_\text{r}$  represents the ratio between the spin wave intensities emitted to the right and to the left. It gradually increases for increasing current density at values beyond that at which vortex dynamics are excited. This trend could be related to the decrease of the emitted wavelength with increasing current which changes the dipolar coupling of the wave to the oscillator, but further studies are needed to quantify this effect.



\newpage

\section{SUMMARY AND CONCLUSIONS}

The continuous development of spintronic devices for information processing technologies aims to design and realize spintronic-based hardware building blocks with large nonlinearity for unconventional computing concepts like neuromorphic computing. Here, we propose a new hybrid device realizing tunable spin-wave emitters which are characterized by merging two subfields, namely the field of spintronic oscillators and the field of nano-magnonics. 
The main advantages of the hybrid magnonic-oscillator system (HMOS) is its nanoscale size, its operation with simple DC currents and its potential to use wave-based effects like frequency-multiplexing to establish efficient connections. We propose a system that utilizes the spatial dynamics of dipolar fields generated by a vortex-based spin transfer oscillator. Coupling these fields to a magnonic waveguide leads to the emission of spin waves. We have verified the functionality of the HMOS via micromagnetic simulations by choosing a set of parameters, including different materials, the element sizes and the gap. It was shown that the HMOS can be used as tunable source for the emission of coherent, propagating spin-waves. This proof-of-concept idea was modeled using simple geometries, no requirement for an external field, and parameters of commonly available materials. Many further modifications can be obtained, e.g., by changing the external magnetic field, the geometric parameters or the static direction of magnetization of the waveguide. Moreover, the wave vector, as well as the frequencies and the amplitudes of the spin waves are tuned by the current density injected into the spin transfer oscillator. In addition, the hybrid system offers the possibility of tuning the direction of emission by the sense of the vortex gyration given by the vortex polarity. The strength of the emission can be changed by the vortex chirality. Combining all these features together, we believe that this device will open a way for the development of new and efficient signal-processing devices for unconventional computing. 

\section*{SUPPLEMENTARY MATERIAL}
See supplementary material for the dynamic magnetic field generated by a vortex at the location of the waveguide for the two cases of polarities.

\begin{acknowledgments}
This research was funded by the European Research Council within the Starting Grant No. 101042439 "CoSpiN" and by the Deutsche Forschungsgemeinschaft (DFG, German Research Foundation) within the Transregional Collaborative Research Center—TRR 173–268565370 “Spin + X” (project B01).
\end{acknowledgments}

\section*{Data Availability Statement}

Data available on request from the authors
The data that support the findings of this study are available from the corresponding author upon reasonable request.

\section*{Conflict of Interest}
The authors have no conflicts to disclose.

\section*{References}

\bibliography{aipsubmission}

\end{document}